# *Akhet Khufu*: archaeo-astronomical hints at a common project of the two main pyramids of Giza, Egypt.

Giulio Magli
Dipartimento di Matematica del  Politecnico di Milano
P.le Leonardo da Vinci 32, 20133 Milano, Italy.
e-mail Giulio.Magli@polimi.it

The architectural complexes composed by the two main pyramids of Giza together with their temples are investigated from an inter-disciplinary point of view, taking into account their astronomical alignments as well as their relationships with the visible landscape. Combining already known facts together with new clues, the work strongly supports the idea that the two complexes were conceived as parts of a common project.

## 1. Introduction

The Giza Plateau overlooks today's city of Cairo, Egypt. Here, in a short period of time (certainly comprised between 2600 and 2450 b.C.) during the so called *Old Kingdom*, the expert workmen and architects of the pharaohs of the 4th Egyptian Dynasty constructed for their rulers two pyramids which are, still today, among the most remarkable achievements of the whole human history. We shall call these pyramids Giza 1 and Giza 2; with side lengths of 230.3 and 215 meters, and heights of 146.6 and 143.5 meters respectively, Giza 1 and Giza 2 are by far greater than the third famous pyramid present on the Plateau, the tomb of the pharaoh Menkaure. This pyramid indeed hardly reaches 65 meters in height and is, therefore, much smaller also of the three pyramids constructed some 80 years before by the pharaoh Snefru in the sites of  Meidum and Dahshur, south of Giza, all of which reach an height of more than 90 meters. The present paper deals only with Giza 1 (the so called *Great Pyramid*) and Giza 2 as well as with their architectural  complexes.
It is very well known that the Giza pyramids were constructed with an extremely high degree of accuracy, in spite of their tremendous building difficulties. In the course of the last two centuries, the accuracy of construction compared to the gigantism of the projects stimulated hundreds of "theories" which have nothing to do with the ingenuity and the way of thinking of the ancient Egyptians as well as the way of working of their architects. Unfortunately, thus, a noisy background of non-scientific theories tends to interfere with any serious approach to the structural, technical and anthropological problems posed by such wonderful monuments. In particular, it is easy to find books (or websites) plenty of strange triangulations, criss-crossing lines or even complex curves traced on maps of the Plateau, which should allegedly represent the hidden legacy of the pyramid builders. On the other side, however, it is clear - at least in the opinion of who writes - that a re-examination of these monuments from an inter-disciplinary point of view would be worth. Such a re-examination would have to take into account, of course, what Egyptology has established in 150 years of research on ancient Egypt, but also, for instance, Geology, Architecture and Structural Engineering.[1] In the present paper, an inter-disciplinary approach is undertaken to study the layout of the Giza 1 and Giza 2 complexes from the point of view of the relationship between architecture, astronomy and landscape. As we shall see, understanding such relationships involves the study of the astronomical orientation of the pyramids, which is so accurate that the rigorous approach of Physics to experimental data must be followed. Another part however, equally important, is connected with the symbolic and religious world of the ancient Egyptians in a more symbolic and "spectacular" way, and for this part we shall make use of simple, rather crude diagrams. Also the reader of the present paper will therefore encounter "images of the Giza plateau with lines overdrawn"; but these lines will not represent any kind of "hidden" legacy; they are indeed based on one thing that everybody knows (or should know) exactly as the ancient Egyptians did, namely, the cycle of the sun during the year.

## 2. The main features of the layouts of the two pyramid's complexes

Each pyramid was the core of a "standard" architectural complex composed by three main elements (see Fig. 1): a "funerary" Temple, located near the pyramid, a "valley" Temple located some hundreds of meters below, near the line of maximal flood of the Nile or near an artificial lake connected to the river, and a straight causeway connecting the two, conceived as a ceremonial road for the Pharaoh's funerals (other annexes, not of interest here, are also present such as smaller "queen" pyramids and boat pits). In the present section the layouts of the Giza 1 and Giza 2 complexes are very briefly described. For full details on the Giza pyramids and their temples the reader is referred to authoritative works such as Petrie (1883), Maragioglio and Rinaldi (1966) and Lehner (1999). [2]

We start from the pyramid complex of Giza 2 which is relatively well preserved. Both the temples are megalithic, with exterior walls made out of enormous limestone blocks weighing up to 250 tons. The funerary temple contained an open rectangular court, while the interior of the Valley Temple presents a characteristic "T" shaped hall; internal halls and courts were pillared with huge granite blocks and cased with huge granite slabs, which are still fully visible in the Valley Temple. The causeway, cut into the rock of the plateau, slopes down straight from the Funerary Temple to a point, which we shall indicate by O', located at the north-west corner of the Valley Temple, reachable from the inside of the building through a spectacular megalithic corridor cased in granite. Over this point passes also the ideal prolongation of the southern side of the pyramid.

The Valley Temple is flanked, on its north side, by another megalithic building called Temple of the Sphinx. The Sphinx itself lies just north of the point O', flanking the causeway. It is a huge statue with the body of a lion and human (or divine) head. The head was carved in a small rocky hill, while the body was obtained excavating a huge precinct from which the blocks for the temple were quarried (Fig. 2). The Sphinx was probably intended to associate the king with the sun god and, as a consequence, the temple in front of it was a sun temple. In any case, all the three monuments are very precisely oriented to due east.

Our knowledge of the layout of the Giza 1 complex is much more incomplete. The funerary temple was dismantled, or perhaps never finished, so that only the floor paved with basalt remains, with huge sockets aimed to held the granite pillars. The causeway starts from these remains and slopes down straight towards the edge of the Plateau, which marks also the boundary between the Giza archaeological zone and the buildings of the modern village of Nazlet el-Saman. At the rocky edge, huge blocks scattered on the escarpment show the point where a monumental ramp once stood, leading the ceremonial road down towards the Valley Temple, which today is lost under the village. For the aims of the present paper, however, it is sufficient to establish that the point (which we shall denote by O) located at the intersection between the ideal prolongation of the northern side of the pyramid and the causeway, and therefore "specular" to the point O', played a special role in the layout of the complex, and the available information appears to be sufficient to confirm this.

To the best of the author's knowledge, only two archaeological campaigns have been carried out with the aim of studying the Giza 1 Valley Temple. The first campaign appears to have been carried out in different phases by Messiha (1983) and by Goyon (1985). The area considered for the soundings is clearly indicated in Goyon's map (Fig. 3) to be at the immediate north-east of point O. The soundings unearthed underlying walls of huge limestone blocks and a narrow corridor cased by one course of basalt blocks. In the interpretation given by these authors, these structures very probably pertained to the Valley Temple, or at least to a very important building, perhaps a sector of a pharaoh's palace. Actually, more recently, an archaeological investigation was operated under the direction of Zahi Hawass (described in Lehner 1999, page 232; see also Note 2) during the construction of sewer channels in the area. In this occasion a basalt pavement probably pertaining to the temple was found, as well as remains of a large settlement of people working at the pyramid's temples; the presence of a pharaoh's palace has also been confirmed (today, only scant remains of the foundation of a building, perhaps a small chapel, can be seen in one of the gardened areas of the traffic islands which are located between the lanes of the main Nazlet el-Saman road).
Accordingly to the map published by Messiha, the causeway ended at the point O, and the building was located in such a way that O occupied (roughly, the map is very inaccurate) its north-west corner. Therefore, this point was in a position with respect to the whole complex similar to that occupied by O' with respect to the Giza 2 complex, and, consequently, the causeway should have had a length comparable to that of Giza 2 which is around 490 meters. According to some 19$^{th}$ century maps and to other authors[3] however, the causeway is shown to continue beyond the zone of the point O for some 350 meters with a sector which bends further north roughly in correspondence of that point, and the temple was located at the very end of this - more than 800 meters long - ceremonial road (such a long causeway would have been in accordance with the description given by Herodotus, who however reported also many other pieces of information most of which are incorrect and/or unclear). Nevertheless, if this was the case, then the presence of the huge 4th dynasty building unearthed by Messiha along the causeway becomes difficult to explain: possibly the further sector of the causeway was a road that once led to the temple from the north-east crossing the inhabited settlement recently discovered. In any case, as mentioned above, for the aims of the present paper what is really important is that the (area of) the point O played a distinguished role in the geometry of the complex, and this is confirmed both by the existence of the building in correspondence of it and by the fact that the road, whether it was still the ceremonial causeway or not, deviated just in correspondence of this building.

**3. Topographical and astronomical references in the pyramid complexes**

It is known that the disposition of the Giza pyramids on the ground is characterized by what is customary called the "Giza diagonal". It is an ideal line which connects the south-east corners of the three pyramids with good accuracy.[4] It was observed already many years ago that this "Giza diagonal" might have had a symbolic meaning, since it points in the direction of the city of Heliopolis, north-east of Giza (according to Jeffreys (1998), the first mention of this fact is contained in a 1852 map by J. Hekekyan). Heliopolis was one of the major cities of ancient Egypt, as well as an extremely important religious center; it was called *Iunu*, pillar, and was a sort of "umbilicus mundi" of the country. The city was sacred to the sun and housed a temple of this god; the sun-priests were skilled in astronomy, to the point that the high priest of Heliopolis was called "Chief of the Observers" already in the times of the 2th dynasty (today, the old city is completely covered by suburbs of Cairo and we know very little about its original extension and structure). A comprehensive study of the inter-visibility between Heliopolis and the pyramids was carried out by Jeffreys (1998). He actually discovered that the sites of the 4th-5th dynasty pyramids whose owners first "declared an affinity" with the sun cult through their monuments and/or their names where chosen in such a way to be visible from Heliopolis (Abu Rawash, Giza, Zawyet El Aryan and the sun temples of Abu Ghurab) (Fig. 4). This is certainly an important "cognitive" aspect of the sacred landscape built in that period; however, at least in the opinion of who writes, *it is not enough* to explain the symbolic meaning of the diagonal alignment at Giza. Actually indeed, as a consequence of the alignment, looking from Heliopolis (and also, of course, from any other point lying nearby the "Giza diagonal" line) the Giza pyramids seem to "contract" on each other and finally their images merge into that of the Great Pyramid. In other words, although the sites of Giza and Heliopolis are intervisible, the effect of the diagonal is that the enormous mass of the second pyramid becomes voluntarily *invisible* from the city. [5]
We will come back on the possible meaning of this "topographical" alignment in the final section, while we shall now proceed to investigate on the astronomical ones. In Fig. 5 a Google-earth satellite image of the area is shown, with three lines emanating from the point O'. The astronomical orientation of these lines is already very well known, and it is related to the cycle of the sun from the spring equinox to the autumn equinox.[6] Indeed:

1) The line O'A' is directed due west. It passes nearby the south side of Giza 2, and therefore the sun at the equinoxes was (and is) seen setting in alignment with the south-east corner of the pyramid in these days.
2) The line O'B' is the alignment defined by the causeway, and it is oriented 14° north of west. The azimuth of the setting sun at the summer solstice at the latitude of Giza is ~28° north of west and therefore this alignment coincides with half-way of the path of the setting sun at the horizon between equinoxes and midsummer (Bauval 1994).
3) The line O'C' is directed towards the midpoint of the segment which separates the south-west corner of Giza 1 and the north-east corner of Giza 2. The azimuth of this line is ~28° north of west, and therefore coincides with that of the sun at the summer solstice. Thus, the midsummer sun is seen setting in between the two pyramids.

The alignments 1) and 3) were rediscovered for the first time by the Egyptologist Mark Lehner during his fieldwork at the Plateau (Lehner 1985b, 1999). He realized that, when the midsummer sun sets, an observer from point O' (or, more generally, from an area in front of the Sphinx) actually witnesses the formation of a spectacular replica of the hieroglyph *Akhet* ⌒ . This is by all means a *hierophany*, a manifestation of divinity which happens every year in dependence of a celestial cycle.[7] Indeed, the hieroglyph Akhet, meaning "horizon", had a profound symbolic meaning for the ancient Egyptians. It was composed by the hieroglyph *djew* ⌣ standing for "primeval mountain" (a mountain with two peaks, "supporting" the heaven) and the sun setting (or rising) in between. The symbol was therefore associated with the afterlife as well, since the solar cycle was associated with life and rebirth.[8]

We now turn to the Giza 1 complex, considering - for the reasons explained in the previous section - the point O as the likely counterpart of the observation point O'. Drawing from O the lines specular to that already drawn from O', we see that they are related to the cycle of the sun from the autumn equinox to the spring equinox. Indeed (Fig. 6):

1) The line OA is directed due west. It passes nearby the north side of Giza 1, and therefore the sun at the equinoxes was (and is) seen setting in alignment with the corner of the pyramid in these days
2) The line OB is the alignment defined by the causeway, and it is oriented 14° south of west. Therefore this alignment coincides with half-way of the path of the setting sun at the horizon between equinoxes and midwinter (Bauval 1994).
3) The line OC is directed towards the Giza 2 complex and passes near the center of the Giza 2 funerary temple, in front of the pyramid. The azimuth of this line is ~28° south of west, and therefore coincides with that of the sun at the winter solstice. Thus, the midwinter sun is seen setting beyond the Giza 2 pyramid.[9]

To the best of the author's knowledge, the alignments 1) and 3) are proposed here for the first time.

**4. Hints at a global project.**

Up to now, I have deliberately avoided to call the Giza 1 and Giza 2 pyramids with the name of the pharaohs which are commonly associated with them. As is well known, however, Giza 1 and Giza 2 are identified with the tombs of the pharaohs Khufu and of his son Khafra (Cheops and Chephren in greek). Due to this attribution, Khafra should be the deified person represented in the face of the Sphinx.

The first key for this attribution of the pyramids is the work of the greek historian Herodotus, who however wrote some two thousands years after their construction. Independent proofs of the attribution are anyhow very clear in the case of Giza 1, because rough workmen's drawings reporting the name of Khufu have been found in four of the chambers located over the main burial chamber; instead, no inscription reporting the name of the builder has ever been found in the Giza 2 pyramid. The attribution of it to Khafra is confirmed by indirect proofs, namely the discovery of several diorite statues (one of them almost intact) of this pharaoh in a pit near the entrance inside the Valley Temple, and the obvious architectural connection between the temples, the Sphinx and the pyramid. Further, it is certain on the basis of inscriptions found in tombs carved around 200 years after the 4th dynasty that the Giza 2 pyramid was attributed to Khafra already at that early time. However, it is *not* certain that the builder of this pyramid was really this pharaoh, since Khafra could have claimed for himself a pre-existing complex.

Actually, some years ago, the present author proposed the possibility of an "inverse chronology" at Giza, namely the idea that Giza 2 could have been built, or at least planned, a *few* years before Giza 1 (Magli 2003, 2005). This proposal was based on a strictly technical analysis of the errors of orientation of the two pyramids from the physical point of view, and will be very briefly recalled here.

The royal pyramids of the fourth dynasty were oriented with a very high, almost maniacal accuracy. Indeed, according to Petrie (1883) and to a detailed study carried out more recently by Dorner (1981), in spite of the bad state of the sides of the buildings today (the casing is lost, except for a few scattered blocks and for the final upper courses of Giza 2) it is possible to determine the deviation from true north with a very high degree of precision, for instance measuring the sockets carved in the rocks for lodging the base blocks. The results of such measures are simply astonishing: Meidum -20.6', Dahshur south -17.3'; Dahshur north -8.7'; Giza 1 -3.4', Giza 2 -6.0', Giza 3 +12.4'. It is certain that such an high accuracy could only be obtained with careful observation of the motion of bright stars, probably circumpolar. However, when the data are reported in a comprehensive plot versus time, they do *not* distribute in a random way inside an error strip, as it would be the case in absence of a time-dependent systematic error. Instead, with the exception of Giza 2, they distribute on an inclined straight line (Fig. 7). It is therefore clear that the method used by the builders to trace the sides of the pyramids was affected by a time-dependent source of error, which can of course be identified with the phenomenon of the precessional motion of the earth's axis. This aimed Spence (2000) to re-investigate the possible methods of orientation used by the ancient Egyptians, searching for a precession-dependent one. She proposed a "simultaneous transit" method which consists in observing the cord connecting two circumpolar stars, namely Kochab and Mizar. When the cord is orthogonal to the horizon, it can be used for precise alignments which however, due to precession, slightly differ from due north and vary with time.[10]

Spence's method accounts very well for the observed variation of the data.[11] For her explanation to work, however, she had to admit that Giza 2 was planned in the opposite season with respect to the other pyramids, so that the stars were in the opposite position with respect to the pole and the "minus" sign of the orientation of Giza 2 could be accounted for. Of course, this is a quite *unsound* explanation since it is rather strange that an important religious procedure such as the foundation of the king's pyramid could occur scattered in time rather then in a fixed period or day. This problem, however, just disappears if the datum for Giza 2 is put before that of Giza 1 in temporal order, and this is the proposal made by the present author (see again Fig. 7).

This hypothesis was sustained by many clues, including the fact that the "best place" for building a pyramid on the Plateau looks rather that of Giza 2, which lies in the higher part of the horizon profile and, at the same time, enjoys of a gentle slope on the east side, which allowed the construction of the causeway without the needs of the huge ramps built for the Giza 1 complex. Further, a geo-morphologic analysis of the Plateau seems to show that the Giza 2 causeway was already existing when the blocks for Giza 1 were quarried (Reader 2001) and, finally, the interpretation of the Sphinx as an image of Khafra is not certain (some Egyptologists, like R. Stadelman, have suggested an attribution to Khufu, others to his son Djedefre). However, to accommodate a inverse chronology at Giza within the well established historical succession of kings, I had to propose that the tomb of Khafra might have originally been the 4th dynasty pyramid of Zawyet El Aryan, which is unfinished and whose attribution is unclear. This was of course a weakness of the theory and, immediately thereafter, Juan Belmonte proposed to give up the idea of an inverse chronology in favour of a common project of the two buildings, an idea on which today we both agree (see Shaltout, Belmonte and Fekri 2007 for further details). According to this proposal, Khufu planned the construction of two pyramids, exactly as his father Snefru did in Dahshur, and later Khafra claimed for himself the one which is slightly smaller.

The hypothesis of a global project at least initially carried out only by Khufu is in agreement with all the abovementioned clues.[12] Further, it does not violate the standard chronology and it is strongly supported by the evidences presented in the previous section, both those which were already known pertaining to the Giza 2 complex and the Giza 1 causeway, as well as the new ones which I am tentatively proposing for Giza 1. All in all, these evidences show that the two complexes have specular alignments with respect to the sides of the pyramids and the causeways, and that each one is embodied with a hierophany at a different solstice, a hierophany which however occurs due to the presence of the other complex.

It might well be that the planners of this gigantic project conceived it also as a sort of calendrical device for the sun cycle, with the Giza 1 complex related to the "southern" part of the yearly movement of the sun, and the Giza 2 complex related to the "northern" one; actually, a likely calendrical interpretation of the Giza 2 complex ad its solar connections have already been proposed (Bauval 2007). Since the motion of the setting sun at the horizon does not occur with constant velocity (it is slower near the solstices and faster near the equinoxes) the azimuths of the causeways do *not* correspond to the intermediate dates between equinoxes and solstices, but to dates which occur more closely to the equinoxes (19 October/21 February and 20 April/19 August respectively). Together with equinoxes and solstices they thus give a sort of *geometrical,* rather than periodical, division of the course of the setting sun during the year, and indeed there is no much evidence of the interest of the Old Kingdom Egyptians for the equinoxes, while of course the cardinal directions (and thus the east-west direction) where fundamental in their symbolic world, as it is shown for instance by the abovementioned orientation of the pyramids. In any case, it is very important to stress that a calendar connection would *not* in any case imply that the temples were used as observatories, because the relative proximity (and the huge masses…) of the "foresights" (the pyramids) make all the alignments cited above rather "symbolic".[13]

In any case, if it is true that the whole complexes where planned according to a common plan, *why* should Khufu have conceived such an ambitious project?

## 5. Discussion

The pathways of the symbolic thought and the feeling of the space as sacred often follow similar patterns, also in cultures which had no contacts whatsoever and were completely disconnected in time, as the fundamental work by Mircea Eliade has authoritatively shown (see e.g. Eliade 1971). Actually there exists a building, constructed 3300 years after the Giza pyramids at thousands of kilometres of distance and by a completely independent culture, where - at least in the opinion of who writes - we can identify similar patterns and find clues for a better understanding of the symbolic mechanisms which possibly motivated the Khufu complex at Giza. Indeed, we find there a similar way of conceiving the constructed landscape and the connection of it with the power and the celestial cycles. It is the so-called *Temple of Inscriptions* of Palenque, in the Yucatan peninsula of Mexico.

The Temple of Inscriptions is the tomb of the great Maya king Pacal, who ruled in the 7th century a.D. The temple, that Linda Schele (1995) calls Pacal's Funerary Mountain, is a huge 9-step pyramid resembling the nine levels of the *Xibalba'*, the Maya after world. A staircase connects the upper level with the tomb; the staircase was filled and closed after burial, but a small conduct runs parallel to it. It is a "psycoduct", aimed to let the soul of the king to reach the living people, in particular the son and new ruler Chan Balam, represented as speaking with the soul of the father in the inscriptions of the temple. The king is buried in a huge sarcophagus whose relief represents him falling down in the after world in the guise of the Maize God. As a dead, he is releasing the power in the hands of his son; however, as a god he is scheduled to reborn, exactly as the renewal of the sun cycle at the winter solstice brings new life to the farming cultures. This symbolic structure of death and renewal and its relationship with the sun cycle was of course known to everybody living in Palenque, and the temple was the inescapable symbol of the king's power to rebirth. However, Pacal wanted also an explicit hierophany to be embodied in the tomb's architecture. Indeed, the temple was oriented in such a way that, as seen from the court and the palace some 100 meters apart, the setting sun at the winter solstice "plunges" into the building as if to enter the underworld through Pacal's tomb, with an angle which is approximately the same as that of the descending stairway (Aveni 1997).

The reason why I find striking similarities with what might have been the global Khufu project at Giza is firstly contained in the title of this paper, *Akhet Khufu*, the horizon of Khufu. Akhet Khufu is the *name* of the Giza 1 pyramid, according to inscriptions present in tombs dated some two hundred years later which report the names of all the three pyramids (Giza 2 at that time was "Khafra is great"). Thus, according to these sources the name of Giza 1 was a precise description of the main hierophany at the site, *a hierophany which however could occur only if Giza 2 existed as well*. The hieroglyphs were actually firstly used with the exact meaning of their images; for instance, the altar for offerings had precisely the same "arrow" form of the hieroglyph standing for "altar". Therefore, it is reasonable to think that the complex was called *Akhet Khufu* because it actually was it: the *Akhet* - the horizon - belonging to Khufu, the king who had "joined the sun-god" as the slightly later (but probably already existing) Pyramid Texts will say (see e.g. Faulkner 1998, utt. 217). If this is true then, as in Palenque, where everybody was aware of the re-birth symbolism of the Pacal tomb, also in the Nile valley everybody knew that the meaning of the *two* giant pyramids was that the king soul was scheduled to live in eternity joined with the sun. Any person looking at the horizon at any time would have been recalled that the horizon in itself *belonged* to Khufu, because exactly this was written as a gigantic hieroglyph, visible from tens of Kilometres away; actually, also nowadays everybody travelling in the Nile valley near Cairo can witness that the horizon, although partially obfuscated by pollution, still belongs to the king who built the unique remaining of the seven wonders of the world (see e.g. Fig. 8).[14] As in Palenque, the re-birth of the king was also embodied explicitly in the architectural layout with a hierophany occurring in this case at the *summer* solstice, since the summer solstice took place roughly in concomitance with the beginning of the Nile flood, essential for the renewal of harvesting cultures in the arid country of Egypt.

Perhaps we shall never reach a definitive proof for this interpretation, and, of course, one could also adopt Spence's explanation for the anomalous orientation of Giza 2 and, further, suppose that it was Khafra to build his pyramidal complex in such a way to realize the *Akhet* hierophany - whose intentional planning appears evident - in honour of his father. However, it remains to explain why this king voluntarily choose the position of his pyramid in such a way that the building becomes invisible from Heliopolis. Instead, if really the aim of Khufu was to claim himself as the *owner of the horizon*, then it makes sense that there was, and still is, only one exception to this otherwise inescapable rule: it is indeed only approaching the city of the Sun god - which was a "symbolic pillar" by itself - that the double-peaked horizon of the great king, slowly and modestly, reduces to a single, although giant, pyramid.

**Acknowledgments**

The author gratefully acknowledges Juan Belmonte for a careful reading of the first version of this manuscript and for many constructive comments.

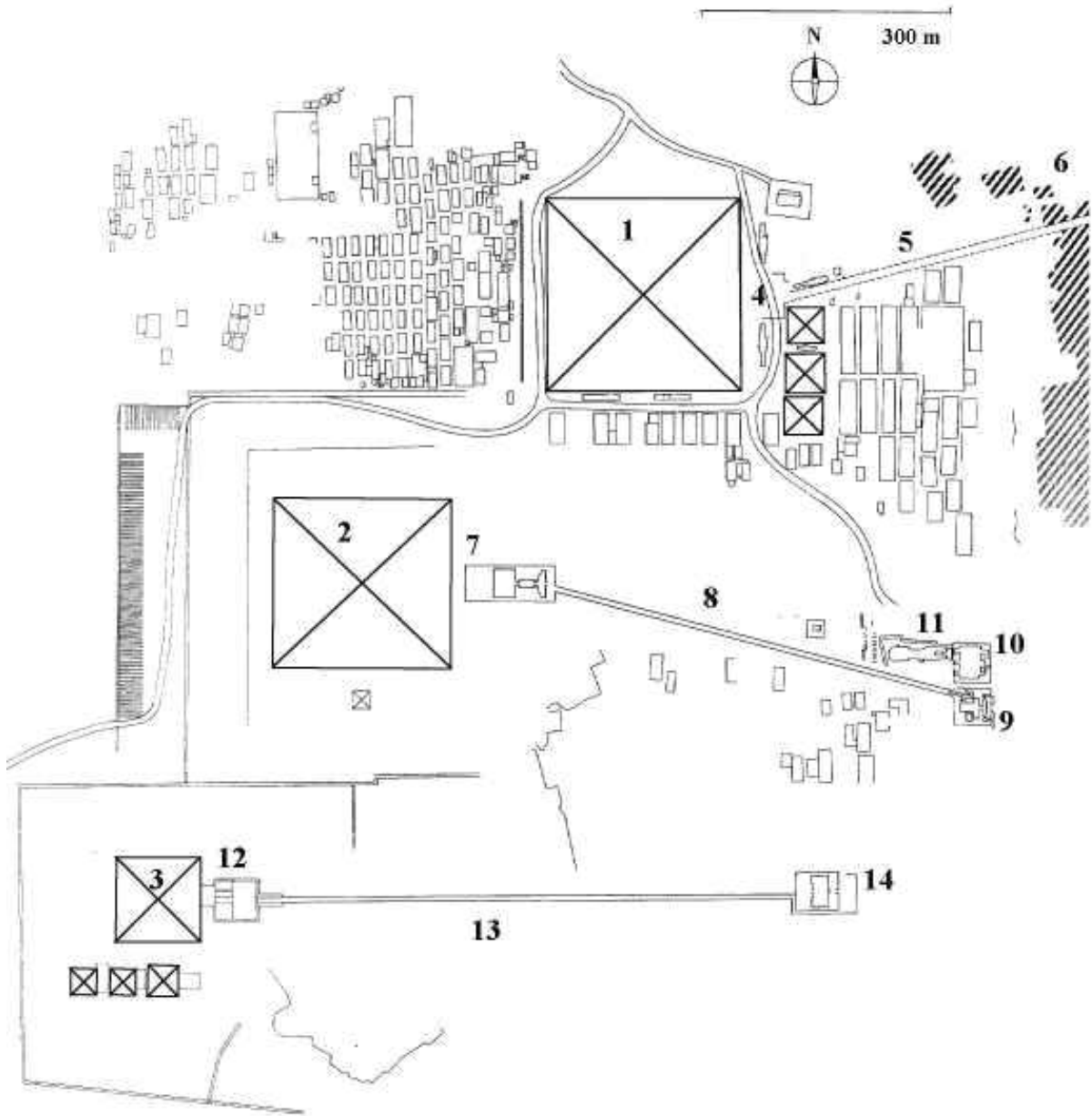

Fig. 1. A schematic map of the Giza Necropolis.
(1-2-3) Main pyramids, (4-7-12) Funerary temples, (5-8-13) Causeways, (6) Modern Village (9-10-11) Giza 2 Valley Temple, Sphinx, Sphinx temple (14) Giza 3 Valley Temple.

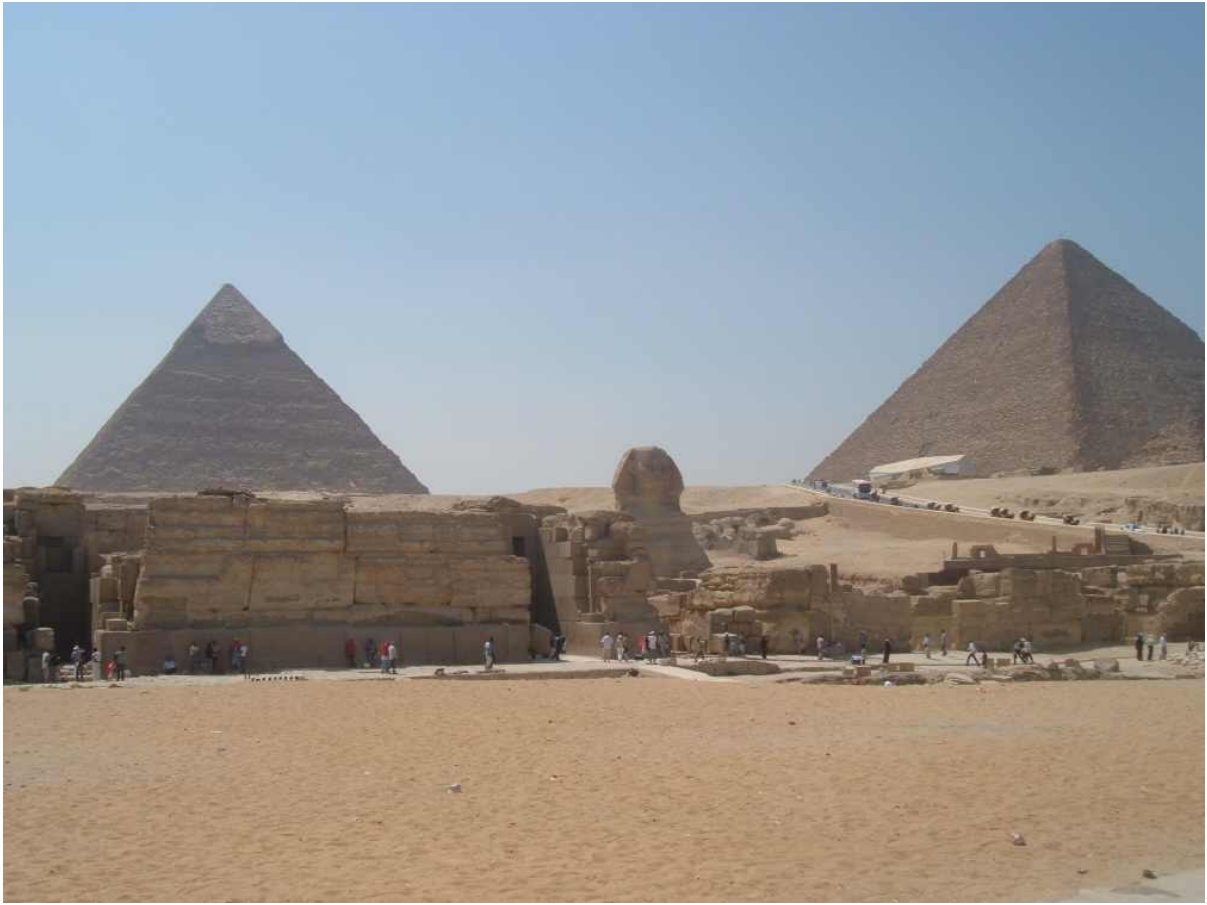

Fig. 2. A view of the Giza 2 Valley complex. From left to right the Valley Temple and the remains of the Sphinx Temple in front of the Sphinx. In the background the Giza 2 and Giza 1 pyramids.

Fig. 3. The area (circled) under the modern village where likely remains of the Giza 1 Valley Temple have been uncovered (adapted from Goyon 1985).

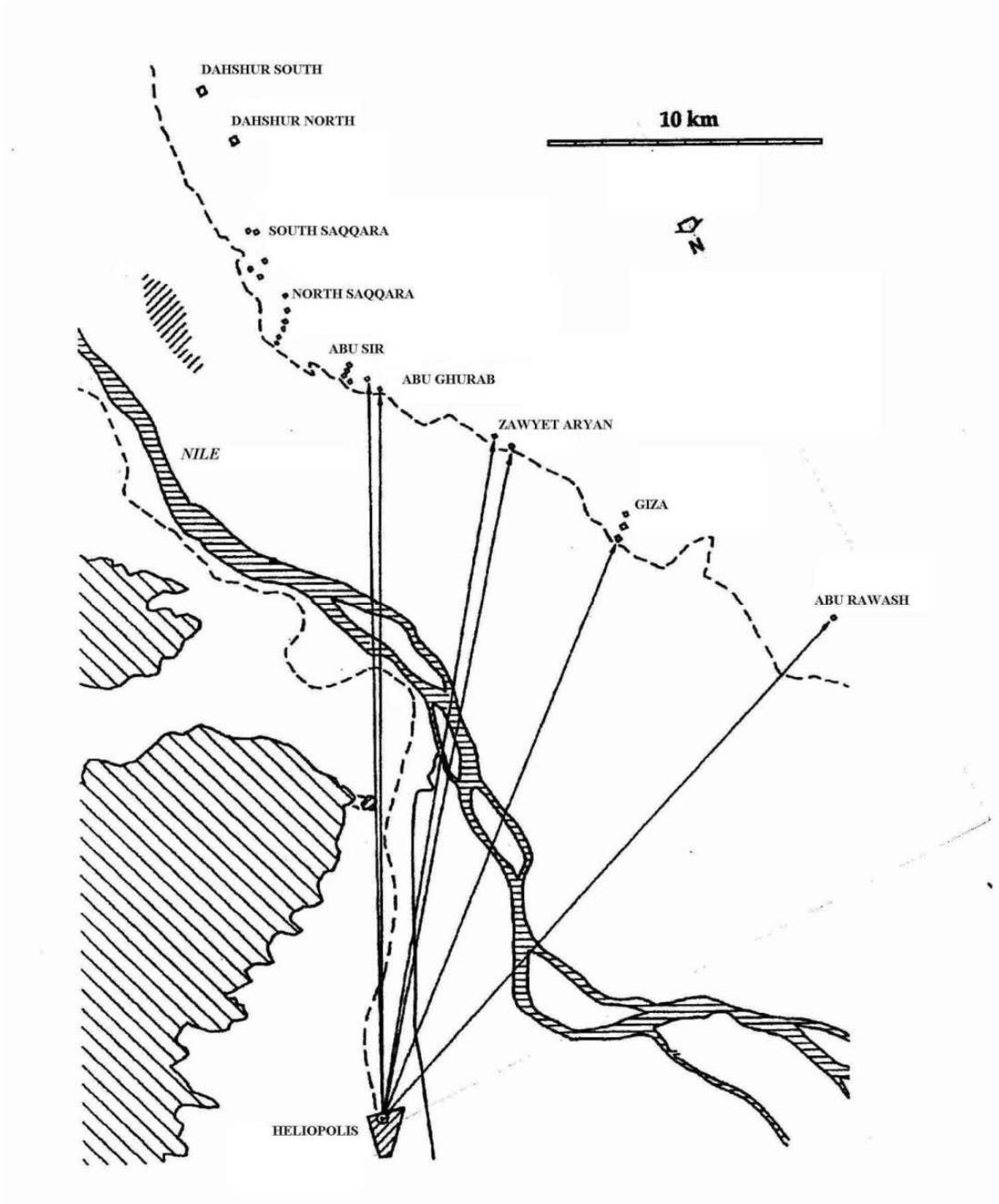

Fig. 4. Inter-visibility between the Old-Kingdom pyramids ad Heliopolis. Straight solid lines indicate mutual intervisibility, which spans from the northernmost pyramid ever constructed, that of Abu Rawash, down to the solar temples at Abu Ghurab. Further south, the view from Heliopolis is blocked by a rocky formation called Muqattam and its western outcrop (where today the Cairo citadel is located). Broken lines indicate the maximal extension of Nile flood (adapted from Jeffreys 1998)

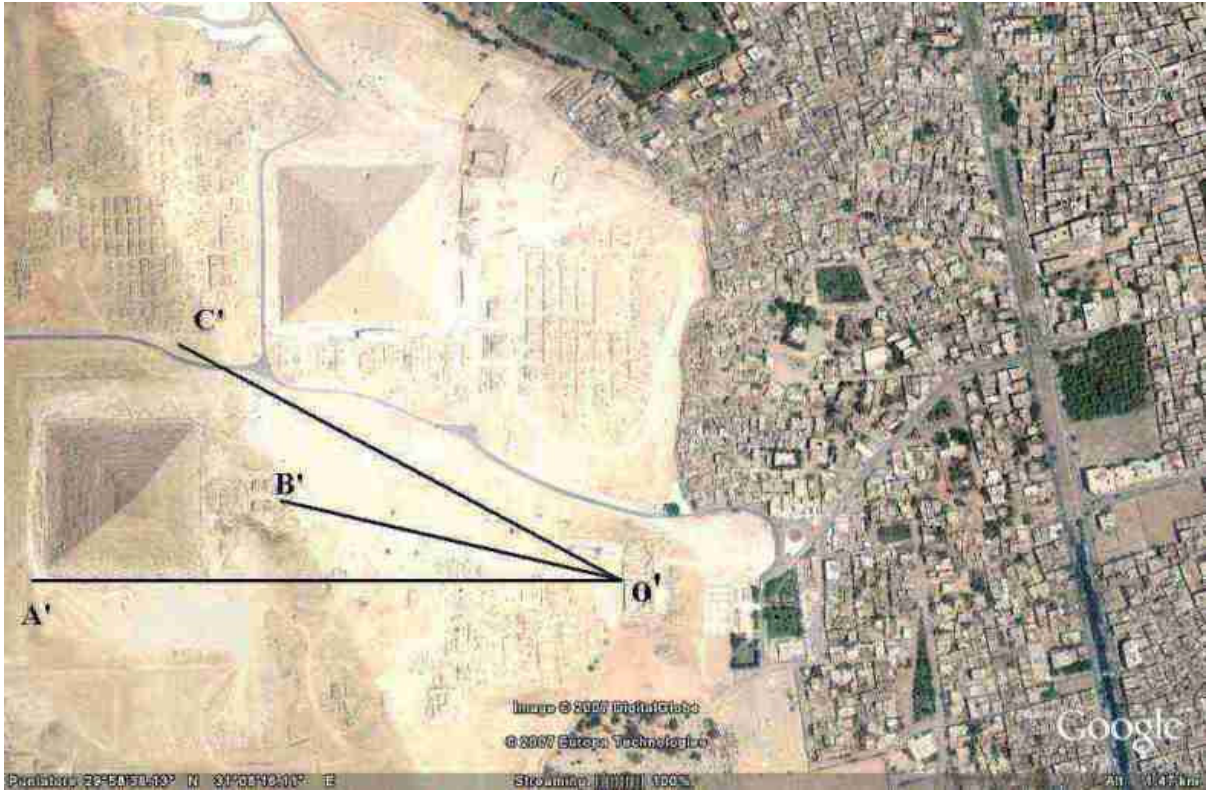

Fig. 5. A Googlearth image of the Giza 1/Giza 2 complexes. See text for discussion.

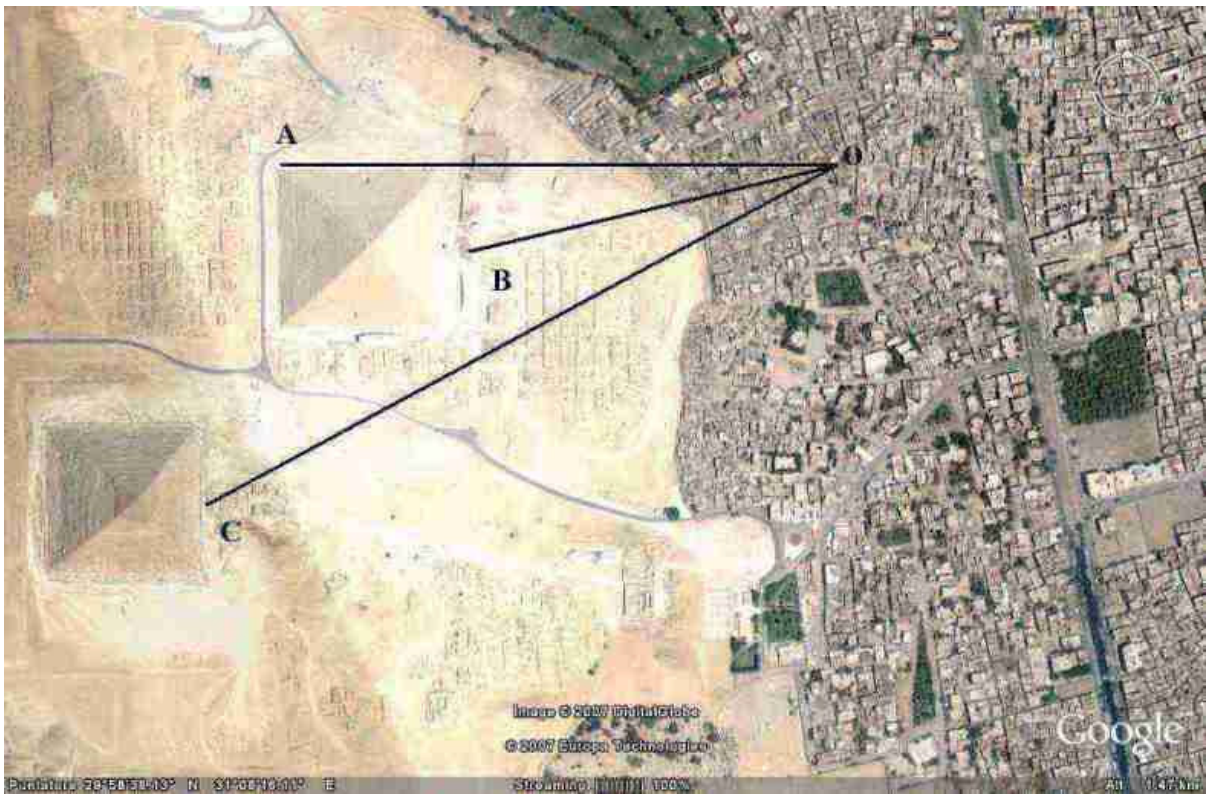

Fig. 6. The same image as in Fig. 3 with different alignments outlined. See text for discussion.

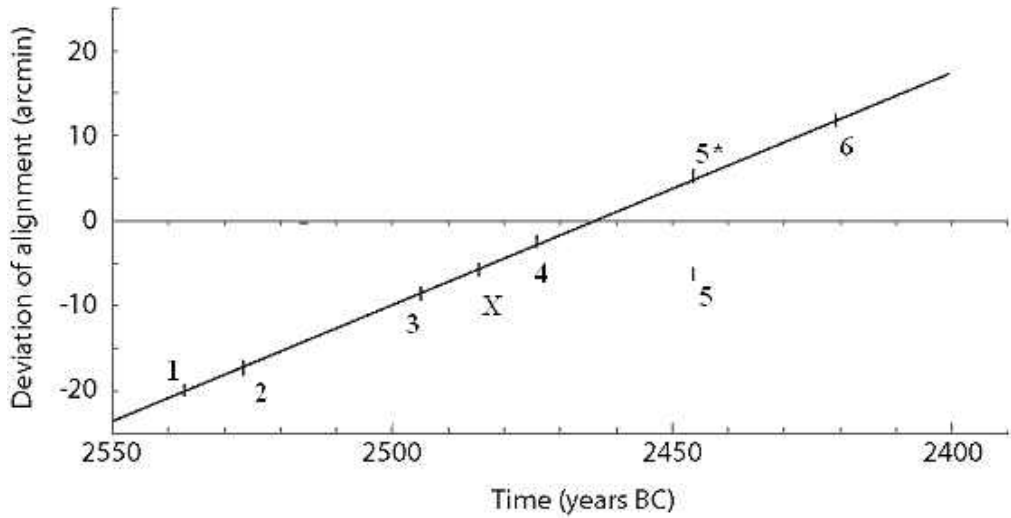

Fig. 7. Deviation from true north (in arc minutes) versus time of the east sides of the 4th dynasty pyramids interpolated (solid line) using the simultaneous transit hypothesis: 1) Meidum 2) Dahshur South 3) Dahshur North 4) Giza 1 5) Giza 2 6) Giza 3. The point 5* indicates the position that the Giza 2 pyramid would have occupied in the case of orientation in the opposite season, while the point X indicates the position that the Giza 2 pyramid would occupy in the case of an inverse chronology. Actually, the points 4 and X are so close that the corresponding dates may overlap, leading to a common project. See text for details.

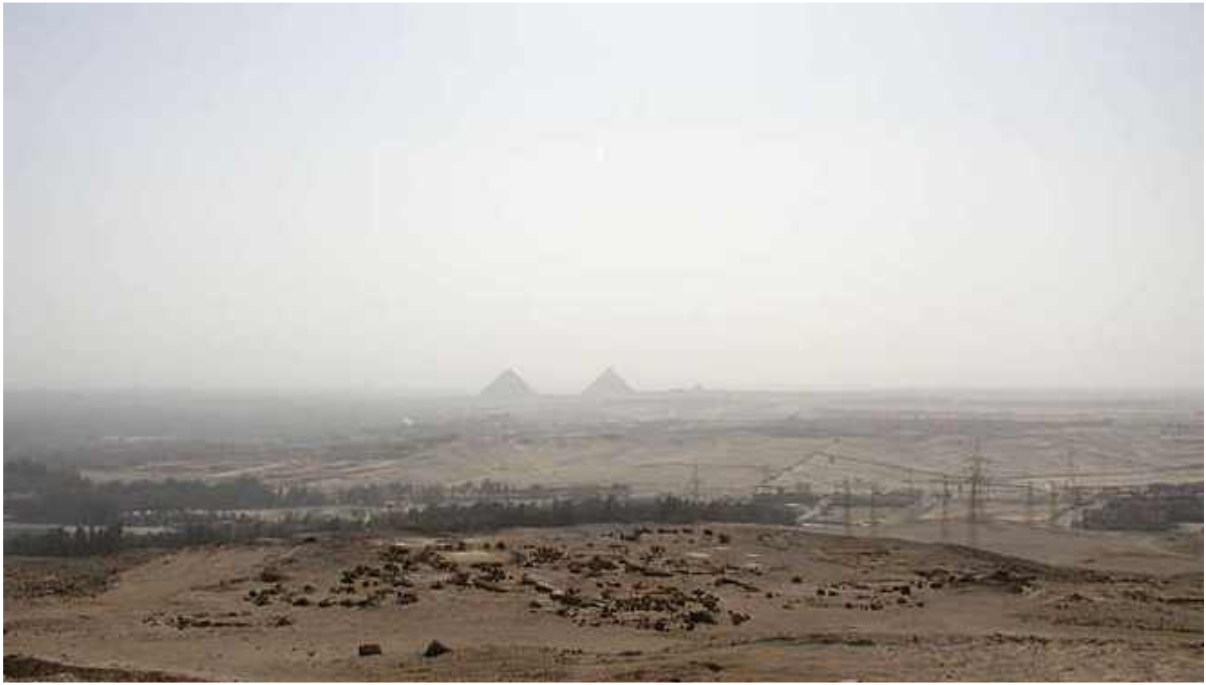

Fig. 8. The view towards Giza from the site of the Abu Rawash pyramid (adapted from www.egyptarchive.co.uk)

[1] It suffices to mention that, still today, the five spaces which lie, one over the other, above the King's chamber of Giza 1 are called "relieving chambers" in spite of the fact that their function *obviously* has nothing to do with the problem of relieving the weight of the pyramid from the ceiling of the room. See Magli (2005) for a complete discussion.

[2] For an account of the most recent discoveries at Giza see Zahi Hawass in http://guardians.net/hawass/pbuildrs.htm

[3] Including Lehner (1985a) and the *Giza Mapping Project* (oi.uchicago.edu/research/projects/giz/comp_model.html)

[4] The corner of the second pyramid "misses" the diagonal by 12 meters (see okadct.googlepages.com/home, unpublished).

[5] The fact that also the third pyramid was aligned to the diagonal remains to be explained as well; further to this, even the "Giza diagonal" is not enough to explain the position of the third pyramid very far into the desert; the discussion of this problem however, and of its possible, although controversial, solution via the so-called "Orion Correlation" theory (Bauval 1989, Bauval and Gilbert 1994) is out of the scopes of present paper.

[6] A slow variation of the ecliptic plane tends to misalign solar references during the millennia. However, as discussed below, all these alignments are "rough" and symbolic; as such they can be witnessed - actually with extraordinary emotions - still today.

[7] Probably the most famous hierophany is that occurring on the Maya-Toltec pyramid called *Castillo* in Chichen Itza', Yucatan, where a light-and-shadow serpent descends the pyramid's stairway at the equinoxes. Exactly as in Giza, this phenomenon was forgotten; it was re-discovered by chance in the 30ts of last century.

[8] The horizon was "protected" by a deity, usually in leonine form. In this respect it may be observed that, in the New Kingdom, the Sphinx was known as Hor-em-akhet, Horus at the Horizon, the god of the rising and setting sun. This god was represented in hieroglyphs as a falcon in the horizon , and it was observed by Wilkinson (1994) that this symbol might have been inspired by the Sphinx as seen in the middle of the two pyramids. Actually, if the hierophany is observed standing in front of the Sphinx, the hieroglyph formed by the sun and the pyramids resembles this last one (Shaltout, Belmonte and Fekri 2007).

[9] Perhaps by chance, this phenomenon recalls the hieroglyph  associated with the glare of the sun rays.

[10] Spence obtained a somewhat late chronology for the 4th dynasty, placing the planning of Giza 1 around 2480 b.C.; an earlier chronology (with the beginning of construction of Giza 1 around the year 2550 b.C.), preferred by most Egyptologists, can anyway be accommodated within the same method supposing that two different stars were used (Belmonte 2001).

[11] The Simultaneous Transit theory was however put in challenge some years ago by a study about another pyramid of the 4th dynasty, that of Djedefre, son of Khufu, at Abu Rawash (Mathieu 2001). In this work it is claimed that the Djedefre pyramid was oriented with an enormous error (of course, enormous from the point of view of the rigid standards of the 4th dynasty, not for today's usual standards) around ¾° . However, the monument was left unfinished and lies in a very bad state, so it is frankly difficult to believe that a *modern* measure of the orientation of this building can be done with an accuracy greater than, say, ½°. Further, both Petrie (1883) and a recent re-evaluation (Shaltout, Belmonte and Fekri 2007) give values much more close to true north at least for the rock cut passage, which is more precisely measurable.

[12] Since the dating of the Giza pyramids has been in the past the subject of several follies, a strictly technical comment is in order here to avoid any kind of misunderstanding. According to the orientation data ordered with the inverse chronology (Fig. 5, point "X"), Giza 2 would have been planned on the ground around 15 years before Giza 1. However, as in any experimental measure, one has to take into account the precision of the instrument used, which in ancient Egypt was, of course, the naked eye, aided with a fork-like viewfinder called *Merkhet* . The precision which is achievable with this kind of instrument by a very expert skywatcher can be estimated to be ±2'(the modern surveying error is instead negligible, being around ±0.2'). This means that the data, taken together with their band of error, have an overlap which allows us to assume safely that the two layouts were actually lid on the ground in the same period.

[13] This holds especially for the alignments of the Giza 1 complex. Indeed, it can be verified by direct inspection that both the Giza 1 and Giza 2 pyramids are visible from the temple area, but it is extremely difficult to verify the accuracy of the alignments due to the presence of the modern buildings and to the fact that the original height of the causeway in the zone of the point O is not known; from this height it depends how much the view at the horizon was originally impeded by the rocky edge of the Plateau.

[14] It is an easy exercise to show that, due to the earth's roundness, the distance in Kilometres at which an object of ``zero`` height can be seen from an height of H meters approximately equals the square root of 13H in kilometres. Thus, for a person 1.70 meters high, the visible horizon is only about 5 km; however if the sight point is not at zero height, the two horizons sum up, and this leads to a theoretical visibility of the two giant pyramids of Giza (considered as 140 meters high) at a distance of more than 47 Kms by a person 1.70 meters high.